\documentclass[12pt]{iopart}
\usepackage{graphicx}
\usepackage{epsf}
\usepackage{epsfig}
\usepackage{iopams}
\newcommand{\fr}{{^F\hspace{-.02in}R}}
\newcommand{\myd}{{\rmd}}
\def\eqref#1{(\ref{#1})}
\begin{document}
\
\title[]{Significance of $c/\sqrt{2}$ in relativistic physics}

\author{C. Chicone\dag, B. Mashhoon\ddag}

\address{\dag\ Department of Mathematics, University of
Missouri-Columbia, Columbia, Missouri 65211, USA}
\address{\ddag\ Department of Physics and
Astronomy, University of Missouri-Columbia, Columbia, Missouri 65211, USA }
\
\ead{MashhoonB@missouri.edu}

\begin{abstract}
In the description of \emph{relative} motion in accelerated
systems and gravitational fields, inertial and tidal accelerations
must be taken into account, respectively. 
These involve a critical speed that in the first
approximation can be simply illustrated in the case of motion
in one dimension. For one-dimensional
motion, such first-order accelerations are multiplied by $(1-V^2/V_c^2)$,
where $V_c=c/\sqrt{2}$ is the critical speed.
If the speed of relative motion exceeds $V_c$, there is a sign reversal
with consequences that are contrary to Newtonian expectations.
\end{abstract}


\maketitle

In four recent papers on the generalized Jacobi equation~\cite{1,2,3,4},
we have considered the consequences of general relativity for the 
\emph{relative} motion of nearby timelike geodesics when the speed
of relative motion is arbitrary but of course less than $c$.
Our main results, which can be clearly seen in the case of 
one-dimensional motion,
depend on whether the speed of relative
motion is above or below the critical speed $V_c=c/\sqrt{2}\approx 0.7 c$.
For one-dimensional motion with
relative velocity $V$, the tidal force acting on a particle
is multiplied by the factor $(1-2V^2/c^2)$, thus
leading to tidal effects for $|V|>c/\sqrt{2}$ that
are counterintuitive when compared to Newtonian expectations.
Specifically, starting from the generalized Jacobi equation (which 
includes only first-order tidal effects) in
a Fermi coordinate system $(T,\bi{X})$ established along the reference
geodesic and restricting attention to a two-dimensional world $(T, Z)$,
the equation of relative motion reduces to 
\begin{equation}
\label{eq0}
\frac{\myd^2 Z}{\myd T^2}+\kappa (1-2 \frac{V^2}{c^2})Z=0,
\end{equation}
where $\kappa(T)=\fr_{TZTZ}$ is the Gaussian curvature of the surface
$(T,Z)$ evaluated along the worldline of the reference geodesic
and $V=\myd Z/\myd T$ (see~\cite{1}).

The purpose of this Letter is to demonstrate that the critical
speed $V_c=c/\sqrt{2}$ appears as well in the physics of translationally
accelerated systems in Minkowski spacetime; that is,
it is a general feature of the theory of relativity. To this end,
we consider two nearby worldlines in Minkowski spacetime: a 
reference worldline $\mathcal{O}$ that is arbitrarily accelerated
and a geodesic worldline $\mathcal{A}$; we are interested
in the motion of the free particle  $\mathcal{A}$ relative
to the noninertial observer $\mathcal{O}$. Referred to the observer's
accelerated system, the motion of $\mathcal{A}$ is 
subject to translational and rotational inertial accelerations.
We will show that relativistic inertial accelerations exist
that go beyond Newtonian mechanics and are essentially due to the
translational  acceleration of $\mathcal{O}$. In this case, the inertial
acceleration of $\mathcal{A}$ in the first approximation 
is multiplied by $(1-2V^2/c^2)$
for motion along the direction of translational acceleration.

Imagine an arbitrary accelerated observer $\mathcal{O}$ following
a worldline $\bar x^\mu(\tau)$ in Minkowski spacetime. 
Here $x^\mu=(t,x,y,z)$ are inertial coordinates in the background global
frame
and $\tau$ is the proper time of the observer, i.e.
$-\myd\tau^2=\eta_{\alpha\beta}\myd\bar{x}^\alpha \myd\bar{x}^\beta$,
where $\eta_{\alpha\beta}$ is the Minkowski metric tensor with
signature $+2$. Unless otherwise specified, we choose units
such that $c=1$. The observer has four-velocity 
$u^\mu=\myd\bar{x}^\mu/\myd\tau$
and translational acceleration $A^\mu(\tau)=\myd u^\mu/\myd\tau$.
Moreover, at each instant of time along its worldline the observer
is endowed with
an orthonormal tetrad frame $\lambda^\mu_{\;\;(\alpha)}$ such that
$\lambda^\mu_{\;\;(0)}=u^\mu$ and 
\begin{equation}\label{eq1}
\eta_{\mu\nu}\lambda^\mu_{\;\;(\alpha)}\lambda^\nu_{\;\;(\beta)}
=\eta_{(\alpha)(\beta)}.
\end{equation}
The variation of this tetrad along the worldline of the observer
is given by
\begin{equation}\label{eq2}
\frac{\myd\lambda^\mu_{\;\;(\alpha)}}{\myd\tau}=
\Phi_{(\alpha)}^{\quad(\beta)} \lambda^\mu_{\;\:(\beta)},
\end{equation}
where
$\Phi_{(\alpha)(\beta)}$ 
is an antisymmetric tensor by equation~\eqref{eq1}.
This \emph{acceleration tensor} has ``electric'' and ``magnetic''
components given respectively by $\bi{a}$ and $\bomega$
in close analogy with the Faraday tensor. That is,
$\Phi_{(0)(i)}=a_i$ and 
$\Phi_{(i)(j)}=\epsilon_{ijk}\,\omega^k$,
where the spacetime scalars $\bi{a}(\tau)$ and $\bomega(\tau)$
represent respectively the local translational acceleration,
$a_i=A_\mu\lambda^\mu_{\;\;(i)}$, and the frequency of rotation
of the local spatial frame of the observer with respect to a local
nonrotating (i.e. Fermi-Walker transported) frame.
We note that 
$\Phi^{\mu\nu}=\Phi^{(\alpha)(\beta)}
  \lambda^\mu_{\;\;(\alpha)}\lambda^\nu_{\;\;(\beta)}$,
hence
\begin{equation}\label{eq2a}
\Phi^{\mu\nu}=A^\mu u^\nu-A^\nu u^\mu
+\epsilon^{\mu\nu\rho\sigma}u_\rho\Omega_\sigma,
\end{equation}
where $\Omega^\sigma=\omega^k\lambda^\sigma_{\;\;(k)}$
and $\epsilon_{0123}:=1$.

Let us now consider the most physically natural (Fermi) coordinate system
in the neighborhood of the worldline of the accelerated
observer~\cite{5s}. That is, we wish to establish a geodesic coordinate
system along the path of the observer based on the tetrad
$\lambda^\mu_{\;\;(\alpha)}$. At any given proper time $\tau$, 
the straight spacelike geodesic lines normal to the observer's
worldline span a Euclidean hyperplane. For a point
on this hyperplane with inertial coordinates $x^\mu$, 
let $X^\mu=(T,\bi{X})$ be the geodesic (Fermi) coordinates such that
$X^0=T=\tau$ and
\begin{equation}\label{eq3}
x^\mu-\bar{x}^\mu(\tau)=X^i\lambda^\mu_{\;\;(i)}(\tau).
\end{equation}
Differentiating this equation and using equation~\eqref{eq2},
we find
\begin{equation}\label{eq4}
\myd x^\mu=\big[P \lambda^\mu_{\;\;(0)}+Q^j 
    \lambda^\mu_{\;\;(j)} \big]\,\myd X^0
+\lambda^\mu_{\;\;(i)}\,\myd X^i,
\end{equation}
where $P$ and $Q$ are given by
\begin{equation}\label{eq5}
P(T,\bi{X})=1+\bi{a}(T)\cdot \bi{X},\qquad 
\bi{Q}(T,\bi{X})=\bomega(T) \times \bi{X}.
\end{equation}
Thus the Minkowski metric 
$\myd s^2=\eta_{\alpha\beta}\,\myd x^\alpha\myd x^\beta $
can now be written in geodesic coordinates as 
$\myd s^2=g_{\mu\nu}\myd X^\mu \myd X^\nu $,
where
\begin{equation}\label{eq6}
g_{00}=-P^2+Q^2, \quad 
g_{0i}=Q_i, \quad
g_{ij}=\delta_{ij}.
\end{equation}
It is simple to see that 
$\det(g_{\mu\nu})=-P^2$ 
and
\begin{equation}\label{eq7}
g^{00}=-\frac{1}{P^2}, \quad 
g^{0i}=\frac{Q_i}{P^2}, \quad
g^{ij}=\delta_{ij}-\frac{Q_iQ_j}{P^2}.
\end{equation}
The observer $\mathcal{O}$ occupies the spatial origin of
the Fermi coordinates, which are admissible for $g_{00}<0$.
In this case, the domain of admissibility has been investigated in~\cite{M}.

The connection coefficients are evaluated using 
equations~\eqref{eq6} and~\eqref{eq7}.  The only nonzero
Christoffel symbols are given by 
\begin{eqnarray}
\label{eq8}
\Gamma^{0}_{00}&=&\frac{\bi{S}\cdot \bi{X}}{P}, \qquad 
\Gamma^{0}_{0i}=\frac{a_i}{P}, \qquad    
\Gamma^{i}_{0j}=-\big(\epsilon_{ijk}\, \omega^k+ \frac{Q_i a_j}{P}\big),\\
\label{eq9}
\Gamma^{i}_{00}&=& P a_i-\frac{\bi{S}\cdot\bi{X}}{P} Q_i
+[\bomega\times (\bomega\times\bi{X})+\dot{\bomega}\times\bi{X}]_i.
\end{eqnarray}
Here an overdot denotes differentiation with respect to $T$
and we have introduced the  vector
\begin{equation}\label{eq10}
\bi{S}(T):=\dot{\bi{a}}+\bi{a}\times \bomega.
\end{equation}
It is important to observe here that $J^\mu:=\myd A^\mu/\myd \tau$
and 
$\Sigma^\mu:=\myd \Omega^\mu/\myd \tau$
are given in the local tetrad frame by $J^{(\alpha)}=(\bi{a}\cdot\bi{a}, \bi{J})$
and $\Sigma^{(\alpha)}=(\bi{a}\cdot\bomega, \dot{\bomega})$,
where $\bi{J}=\dot{\bi{a}}+\bomega\times \bi{a}$.

The free particle $\mathcal{A}$ follows a geodesic
in the new coordinate system. To express its motion
relative to the reference observer $\mathcal{O}$, 
we need the reduced geodesic equation~\cite{1} in the
new coordinate system
\begin{equation}\label{eq11}
\frac{\myd^2 X^i}{\myd T^2}-(\Gamma^0_{\alpha\beta}\,
  \frac{\myd  x^\alpha}{\myd T}\frac{\myd  x^\beta}{\myd T})
\frac{\myd  X^i}{\myd T}
+\Gamma^i_{\alpha\beta}\,\frac{\myd  X^\alpha}{\myd T}
\frac{\myd  X^\beta}{\myd T}=0,
\end{equation}
where the Christoffel symbols are given by equations~\eqref{eq8}
and~\eqref{eq9}. The result is 
\begin{equation}\label{eq12}
\fl\frac{\myd^2 \bi{X}}{\myd T^2}+2\bomega\times\bi{V}
  +\bomega\times(\bomega\times\bi{X})+\dot{\bomega}\times\bi{X}
 +P\bi{a}-\frac{1}{c^2P} (\bi{Q}+\bi{V})(\bi{S}\cdot\bi{X}+2\bi{a}\cdot\bi{V})=0,
\end{equation}
where $\bi{V}=\dot\bi{X}$ and $P=1+\bi{a}\cdot \bi{X}/c^2$.
Here the presence of the speed of light $c$ has been made explicit so 
that relativistic corrections to the Newtonian inertial accelerations can be 
easily identified.

Equation~\eqref{eq12} contains the fully relativistic inertial accelerations
of a free particle with respect to an accelerating and rotating reference
system and has been derived in various forms by a number of authors
(see~\cite{8,9,10,11,11b,11c} and the references cited therein).
In the case of pure rotation ($\bi{a}=0$), the inertial accelerations
in equation~\eqref{eq12} are just as in Newtonian mechanics. 
We note that a critical speed of $c/\sqrt{2}$ has appeared in the treatment
of the motion of relativistic charged particles in the field of rotating
magnetic lines of force corresponding to pulsar magnetospheres~\cite{5,6,7}.
This critical speed turns out to be due to the
particular mechanical model of the electromagnetic system discussed in~\cite{5}.
In the theory of relativity, there is no critical speed associated with rotation
per se as is evident from equation~\eqref{eq12}. The situation is different
in the case of translational acceleration; however, 
before we turn to the case of purely translational accelerations, 
let us note the existence of relativistic inertial accelerations 
in equation~\eqref{eq12} that are due to the coupling of acceleration and rotation.

\begin{figure}
\vspace{1in}
\centerline{\psfig{file=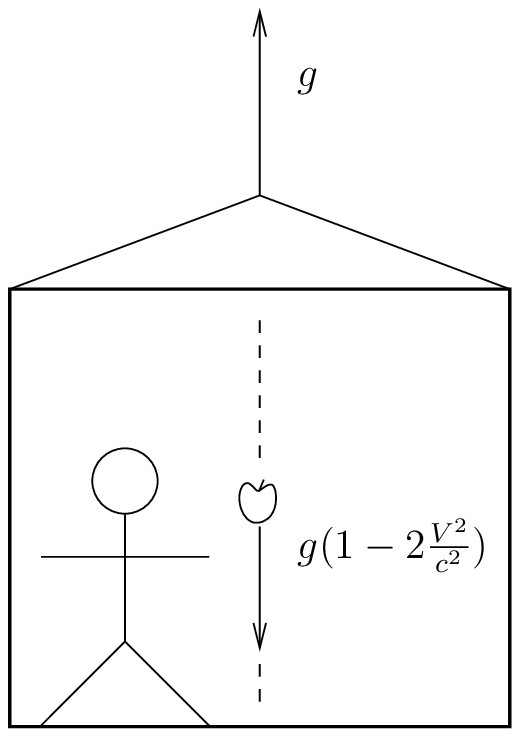,width=15pc}\qquad\psfig{file=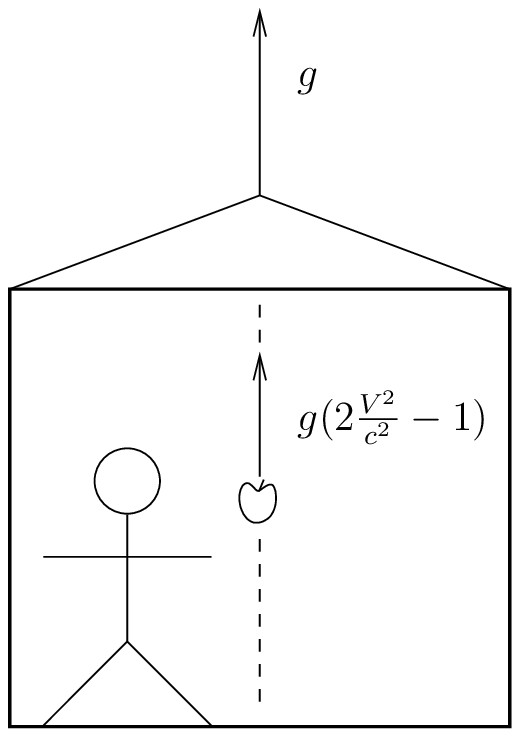,width=15pc}}
\caption{The inertial acceleration of a free particle according to the
accelerated observer. In the left panel $|V| <c/\sqrt{2} $
and in the right panel $|V| > c/\sqrt{2} $.  \label{fig:1}}
\end{figure}
Set $\bomega=0$ and note that equation~\eqref{eq12} reduces to 
\begin{equation}\label{eq13}
\frac{\myd^2 \bi{X}}{\myd T^2}+P\bi{a}
-\frac{\bi{V}}{c^2 P}(\dot{\bi{a}}\cdot\bi{X}+2\bi{a}\cdot\bi{V})=0.
\end{equation}
In the Newtonian limit $(c\to\infty)$, this equation reduces to a
standard result: from the viewpoint of the observer, the
free particle has acceleration $-\bi{a}$. 
Writing  $\bi{a}$ in equation~\eqref{eq13} in terms
of its components parallel and perpendicular to the 
instantaneous direction of the velocity
$\bi{V}=V\widehat{\bi{V}}$,
$\bi{a}=a_\parallel\widehat{\bi{V}}+\bi{a}_\bot$,
the parallel component of the inertial acceleration in
equation~\eqref{eq13} to first order in $\bi{a}$ and neglecting
$\dot{\bi{a}}$ 
is given by
$-a_\parallel(1-2 V^2/c^2)\widehat{\bi{V}}$.
To illustrate this result, it is useful to assume one-dimensional motion
of $\mathcal{A}$ 
along the $Z$-direction such that $\bi{a}=a(T)\widehat{\bi{Z}}$.
In this case, to first order in $a$ and neglecting $\dot a$, 
the inertial acceleration is simply given by 
$-a(1-2 V^2/c^2)$. Without these approximations, 
the acceleration of the free particle relative to the observer
at its position ($\bi{X}=0$) is 
\begin{equation}\label{eq14}
\frac{\myd^2 Z^2}{\myd T^2}\Big|_{Z=0}=-a(1-2\frac{V^2}{c^2}),
\end{equation}
so that if $|V|< V_c=c/\sqrt{2}$, the free particle has the expected
direction of
acceleration as observed by $\mathcal{O}$. On the other hand,
for $|V|> V_c$ the inertial acceleration of $\mathcal{A}$ reverses direction.
This situation is schematically illustrated in Figure~\ref{fig:1}.
If $|V|=V_c$, $\mathcal{A}$ has no inertial acceleration at $Z=0$~\cite{11}.

Let us recall that in the standard discussion of Einstein's heuristic
Principle of Equivalence, the observer at rest in the ``elevator'' that
is accelerated with acceleration $\bi{a}$ determines
the acceleration of a free particle $\mathcal{A}$ (``apple'') that moves
past the observer along the same line. 
The standard discussion is limited to the Newtonian limit
and in that limit the acceleration of $\mathcal{A}$ is $-\bi{a}$.
However, if relativistic effects are taken into account, then the
correct answer is given by equation~\eqref{eq14}. 
For relative speed above $V_c=c/\sqrt{2}$, the
direction of acceleration is opposite to the result of Newtonian mechanics,
which is counterintuitive, since our intuition is based
on Newtonian expectations.

To clarify these issues further, let us consider a uniformly accelerated
observer in hyperbolic motion with acceleration $g>0$ moving along
the positive $z$-direction in the background global inertial frame. 
The worldline of the observer is given by 
\begin{equation}\label{eq15}
\bar{t}=\frac{1}{g}\sinh g\tau,\qquad
\bar{x}=\bar{y}=0, \qquad
\bar{z}=z_0+\frac{1}{g}(-1+\cosh g\tau),
\end{equation}
so that at $\bar{t}=\tau=0$, the observer is at rest at $\bar{z}=z_0$.
The natural nonrotating tetrad frame along the observer's worldline
has nonzero components
$\lambda^0_{\;\;(0)}=\lambda^3_{\;\;(3)}=\gamma$, 
$\lambda^3_{\;\;(0)}=\lambda^0_{\;\;(3)}=\beta\gamma$
and
$\lambda^1_{\;\;(1)}=\lambda^2_{\;\;(2)}=1$,
where $\beta=\tanh g\tau$ and $\gamma=\cosh g\tau$. It follows
from equation~\eqref{eq3} that the inertial coordinates are related
to the coordinates of the accelerated (geodesic) frame via
\begin{eqnarray}
\label{eq16} t&=& (Z+\frac{1}{g})\sinh g T,\\
\label{eq17} x&=& X, \quad y=Y,\\
\label{eq18} z&=&z_0-\frac{1}{g}+(Z+\frac{1}{g})\cosh gT , 
\end{eqnarray}
so that at $X^\mu=(\tau,\mathbf{0})$, equations~\eqref{eq16}--\eqref{eq18}
reduce to equation~\eqref{eq15}. The new (Rindler) coordinates 
are $T, X, Y\in (-\infty,\infty)$ and $Z\in (-1/g, \infty)$, where $Z = - 1/g$ is the Rindler
horizon and corresponds to a null cone in the inertial frame, 
since $(Z + 1/g )^2 =( z - z_0 + 1/g )^2 - t^2$.

The free particle follows a straight line in the inertial frame
and this fact can be used to find an explicit solution 
 to equation~\eqref{eq13}
in this case.
We limit our discussion to motion of the free particle along the $z$-direction
\begin{equation}
\label{eq19} 
z=v_0t+z_0,
\end{equation}
such that at $t=0$ it passes the observer with relative speed $v_0$. 
It follows from
equations~\eqref{eq16}--\eqref{eq18} that
\begin{equation}
\label{eq20} 
Z=\frac{1}{g}\Big(-1+\frac{1}{\cosh gT-v_0\sinh gT}\Big).
\end{equation}
Computing the inertial acceleration of the free particle with respect
to the observer, we find that $\ddot Z(T=0)=-g(1-2v_0^2)$,
where $v_0=\dot Z(T=0)$. This is the same result that we obtained before
in equation~\eqref{eq14} for the general case of variable acceleration.

It is straightforward to extend our analysis of the motion of $\mathcal{A}$
 relative
to $\mathcal{O}$ to the curved spacetime of general relativity along the lines
indicated in~\cite{ 12, 9, 10}.  The origin of the interesting factor 
 $( 1 - 2 V^2/c^2 )$ turns out to be the same for both tidal accelerations 
as well as the
translational inertial acceleration: it comes about in the transition from the
standard geodesic equation to its reduced form~\eqref{eq11}.  
That is, the factor $( 1 -2 V^2/ c^2 )$ is basically due to 
the representation of the motion of $\mathcal{A}$
in terms of the proper time of the observer $\mathcal{O}$ rather 
than the proper time
of the free particle $\mathcal{A}$.

The physical phenomena associated with the factor $(1 - 2 V^2/c^2)$
in equations \eqref{eq13} and \eqref{eq14} come about only when 
the motion of $\mathcal{A}$ is referred to Fermi coordinates; 
that is, they do not 
in general appear in any other coordinate system.
Nevertheless, 
equations \eqref{eq13} and \eqref{eq14} are constructed from 
scalar invariants and thus express real physically measurable effects. 
This circumstance may be illustrated as follows: 
the magnitude of the solution of equation~\eqref{eq13}, $|{\bf X}(T)|$, 
is the proper distance between the worldlines of $\mathcal{O}$ and 
$\mathcal{A}$, measured along a spacelike geodesic that is normal 
to the worldline of the observer $\mathcal{O}$ at its proper time $T$. 
As such, this invariant quantity can be computed in any admissible 
system of coordinates. Fermi coordinates are advantageous in 
practice as they are the most physically natural system of coordinates; 
therefore, equations~\eqref{eq13} and~\eqref{eq14} can be subjected 
to \emph{direct} 
experimental test if the observer employs  
a local coordinate system that closely approximates a Fermi system.

Finally, let us remark that the proper critical speed
$c/\sqrt{2}$ has also been discussed for
one-dimensional relative motion along the radial direction
in the context of the exterior Schwarzschild
geometry in~\cite{13}.

\section*{Acknowledgment}
BM is grateful to Tinatin Kahniashvili for helpful correspondence.

\end{document}